\documentclass[pre,numerical,reprint]{revtex4-1}
\usepackage[english]{babel}
\usepackage{microtype}
\usepackage{amsmath,amsthm,amssymb,amsfonts}
\usepackage{mathtools}
\usepackage{siunitx}
\usepackage{tabularx,graphicx}
\usepackage{booktabs}
\usepackage{xr}
\usepackage[breaklinks=true,colorlinks=true,linkcolor=blue,urlcolor=blue,citecolor=blue]{hyperref}

\usepackage[bordercolor=none,backgroundcolor=none,textsize=scriptsize]{todonotes}
\newcommand{\ddf}[1]{\mathrm{d} #1 \,}
\DeclarePairedDelimiter\avg{\langle}{\rangle}

\DeclareMathOperator\const{const.}
\newcommand{\e}{\mathrm{e}}

\newcommand{\kB}{k_{\text{B}}}

\newcommand{\Birr}{B_{\text{irr}}}
\newcommand{\Bqe}{B_{\text{qe}}}
\newcommand{\Binst}{B_{\text{inst}}}

\newcommand{\pirr}{p_{\text{irr}}}
\newcommand{\pqe}{p_{\text{qe}}}
\newcommand{\pinst}{p_{\text{inst}}}

\newcommand{\koff}{k_{\text{off}}}
\newcommand{\kon}{k_{\text{on}}}

\newcommand{\Noff}{N_{\text{off}}}

\newcommand{\Nbb}{N_{b \to b}}
\newcommand{\Nbu}{N_{b \to u}}
\newcommand{\Nub}{N_{u \to b}}
\newcommand{\Nuu}{N_{u \to u}}

\newcommand{\Fmode}{F_{\text{mode}}}

\newcommand{\Fbb}{F^{b \to b}}
\newcommand{\Fbu}{F^{b \to u}}
\newcommand{\Fub}{F^{u \to b}}
\newcommand{\Fuu}{F^{u \to u}}

\newcommand{\dxoff}{\Delta x_{\text{off}}}
\newcommand{\dxon}{\Delta x_{\text{on}}}

\newcommand{\toff}{t^{\text{off}}}
\newcommand{\ton}{t^{\text{on}}}

\newcommand{\var}{\operatorname{var}}

\DeclareSIUnit\molar{\textsc{M}}

\begin{document}

\preprint{}

\title{Rebinding kinetics from single-molecule force spectroscopy experiments close to equilibrium}

\author{Jakob T\'{o}mas Bullerjahn}
\affiliation{Department of Theoretical Biophysics, Max Planck Institute of Biophysics, 60438 Frankfurt am Main, Germany}
\author{Gerhard~Hummer}
\email{gerhard.hummer@biophys.mpg.de}
\affiliation{Department of Theoretical Biophysics, Max Planck Institute of Biophysics, 60438 Frankfurt am Main, Germany}
\affiliation{Institute of Biophysics, Goethe University Frankfurt, 60438 Frankfurt am Main, Germany}
\date{\today}

\begin{abstract}
Analysis of bond rupture data from single-molecule force spectroscopy experiments commonly relies on the strong assumption that the bond dissociation process is irreversible.  However, with increased spatiotemporal resolution of instruments it is now possible to observe multiple unbinding-rebinding events in a single pulling experiment.  Here, we augment the theory of force-induced unbinding by explicitly taking into account rebinding kinetics, and provide approximate analytic solutions of the resulting rate equations.  Furthermore, we use a short-time expansion of the exact kinetics to construct numerically efficient maximum likelihood estimators for the parameters of the force-dependent unbinding and rebinding rates, which pair well with and complement established methods, such as the analysis of rate maps.  We provide an open-source implementation of the theory, evaluated for Bell-like rates, which we apply to synthetic data generated by a Gillespie stochastic simulation algorithm for time-dependent rates.  
\end{abstract}

\maketitle

\section{Introduction}

Single-molecule force spectroscopy (SMFS) is an experimental technique used to gauge the stability of intermolecular and intramolecular bonds by repeatedly stretching and breaking them via external loading protocols~\cite{Noy2011}.  Among other applications, SMFS has been used to study (poly)protein unfolding~\cite{RiefGautel1997, SchlierfLi2004, HoffmannDougan2012, RicoGonzalez2013}, and to explore the mechanical strength of cell adhesion molecules~\cite{MarshallLong2003, HeleniusHeisenberg2008, BoyeLigezowska2013} and ligand-receptor complexes~\cite{MerkelNassoy1999, StahlNash2012, RicoRussek2019}.  The yield forces measured in force spectroscopy experiments are conventionally analyzed with the help of minimal theories~\cite{EvansRitchie1997, HummerSzabo2003, DudkoHummer2006, BullerjahnSturm2014, AdhikariBeach2020}, which have been remarkably successful despite the enormous reduction in complexity from a high-dimensional molecular system to a low-dimensional kinetic or diffusive representation~\cite{Makarov2016}.  These theoretical models provide analytic expressions for experimentally accessible observables, such as the mean lifetime of a bond or the distribution of rupture forces measured under force-clamp and force-ramp protocols, respectively.  However, most of these theories model the unbinding process as an \emph{irreversible} transition, which is only a reasonable assumption to make for large pulling speeds and for molecules with slow rebinding or refolding rates.  Furthermore, technical advances have in recent years led to microsecond response times in force actuators~\cite{RicoGonzalez2013, NeupaneFoster2016, YuSiewny2017} that paved the way for elaborate transition-path measurements~\cite{NeupaneFoster2016, NeupaneHoffer2018} and the detection of rapid unfolding and refolding transitions in local regions of macromolecules~\cite{YuJacobson2020}.  One therefore must take into account rebinding effects when analyzing data recorded at such high time resolution.  

Up until now, models of force-induced rupture have only addressed the problem of rebinding indirectly.  Friddle \emph{et al.}~\cite{FriddleNoy2012} considered the extreme case of having unresolved rebinding events below a certain threshold force.  For forces above this threshold, Friddle \emph{et al.} assumed rebinding to be negligible.  The resulting mean rupture force, as a function of the loading rate $\dot{F} = \mathrm{d}F/\mathrm{d}t$, then has a nonzero vertical intercept at a value corresponding to the threshold force.  Friddle followed up on this idea in Ref.~\onlinecite{Friddle2016}, where he constructed an approximate closed-form solution to the coupled rate equations describing the unbinding-rebinding kinetics of a two-state system.  

In this paper, we revisit the kinetic modeling of reversible bond rupture and rebinding under a time-dependent force.  Force-ramp protocols have significant advantages over their force-clamp counterparts.  In particular, by sweeping through a wide range of forces, they are essentially guaranteed to reveal state transitions~\cite{JunkerZiegler2009}.  However, the application of time-dependent forces complicates the kinetic modeling.  Here we address this main drawback of slow force-ramp experiments compared to force-clamp experiments.  

Starting from the rate equations of Friddle~\cite{Friddle2016}, we extend the theory to the regime of small force loading rates, where rebinding takes place.  In this regime, nonequilibrium effects are suppressed and multiple transition events can occur in a single pulling trace.  We show that a quasi-equilibrium treatment gives excellent results in this limit.  To facilitate the analysis of experimental data, we provide an asymptotically exact maximum likelihood estimator for the parameters of Bell-like rates to complement the established nonparametric method of rate-map analysis~\cite{DudkoHummer2008, ZhangDudko2013}.  

The paper is structured as follows.  In Sec.~\ref{sec:general-theory}, we briefly review the two-state Markov modeling ansatz behind the coupled rate equations, present their formal solution, and provide general recipes for constructing the corresponding rupture force distributions (RFDs) and calculating the average number of unbinding events, as well as its higher moments.  Section~\ref{sec:approximations} discusses closed-form approximations to the formal solution of the rate equations, in particular Friddle's instantaneous rebinding approximation, as well as our quasi-equilibrium approximation.  We then proceed to develop a likelihood function in Sec.~\ref{sec:data-fitting} that can be maximized to give estimates of model parameters and used to analyze time series involving unbinding-rebinding kinetics driven by force.  The resulting estimators become exact in the limit of infinitesimally small time steps and are therefore not limited to quasi-equilibrium situations.  To demonstrate the applicability of our approach, we use it to analyze simulation data generated by a modified Gillespie stochastic simulation algorithm~\cite{ThanhPriami2015} that accounts for the time-dependence of the rates.  Finally, we conclude in Sec.~\ref{sec:conclusions} with a summary of our results.  A computationally efficient open-source implementation of the theory, as well as the simulation code, are provided in a data analysis package~\cite{JuliaScript} written in Julia~\cite{BezansonEdelman2017}.

\section{General theory}\label{sec:general-theory}

Conformational changes of macromolecules or the unbinding-rebinding kinetics of a protein-ligand complex can, to a first approximation, often be described as a two-state Markov process, characterized by the following coupled rate equations:
\begin{equation}\label{eq:rate-equations}
\begin{aligned}
\dot{B}(t) & = - \koff(t) B(t) + \kon(t) U(t) \, ,
\\
\dot{U}(t) & = - \kon(t) U(t) + \koff(t) B(t) \, .  
\end{aligned}
\end{equation}
Here, a dot indicates the time derivative, \emph{i.e.}, $\smash{\dot{B}(t)} = \mathrm{d}B(t) / \mathrm{d} t$; $B(t)$ and $U(t)$ are the relative populations inside the bound and the unbound states, respectively; and $k_{\text{off/on}}(t)$ denote the time-dependent unbinding (``off'') and rebinding (``on'') rates, respectively, driving the kinetics.  The time dependence of the rates is assumed to originate solely from an external force protocol $F(t)$, ignoring the thermal fluctuations inherent in molecular systems~\cite{HummerSzabo2003}.  Under the assumption that $F(t)$ is monotonic (\emph{i.e.}, an invertible function for all $t$), we can therefore freely switch between the variables $t$ and $F$ with the help of the loading rate $\smash{\dot{F}[F]} = \mathrm{d}F / \mathrm{d}t$.  We shall reserve parentheses for time-domain functions $f(t)$ and denote force-dependence via the bracket notation
\begin{equation*}
f[F] \equiv f \big( t(F) \big) \, .  
\end{equation*}

\begin{figure}[t!]
\begin{center}
\includegraphics{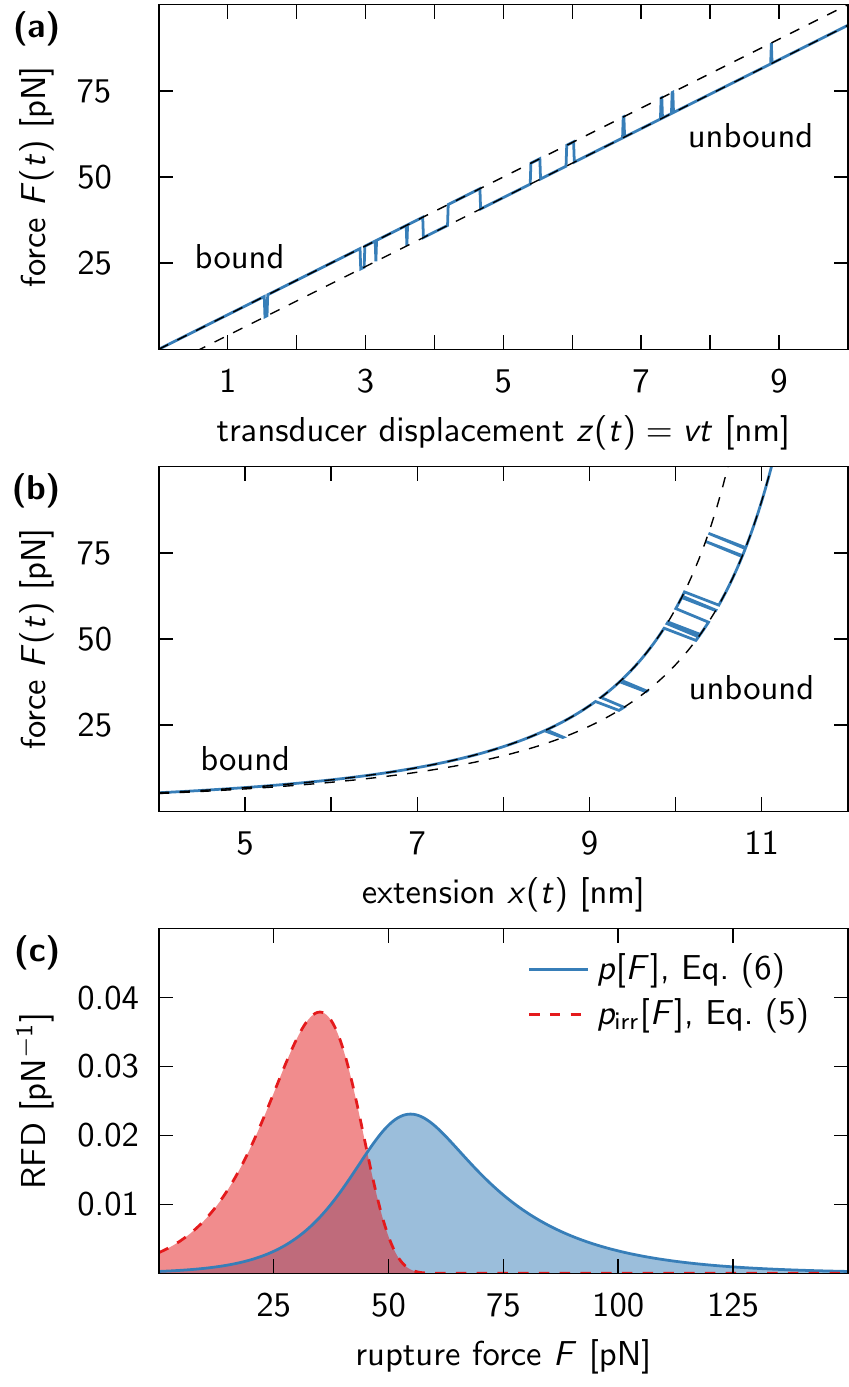}
\caption{Schematic of SMFS experiments using dynamic force protocols.  (a)~Idealized force-extension curve for a linear force protocol, defined by a constant loading rate $\smash{\dot{F}} = \kappa v$ and a state-dependent force offset $F^{*} = \kappa y(s)$.  Here, $\kappa$ and $v$ denote the spring constant and speed of the force transducer, respectively.  (b)~More realistic force-extension curve, where the bound ($s=b$) and unbound ($s=u$) force branches are modeled by wormlike chains with different contour lengths.  In this case, the loading rate $\smash{\dot{F}[F]}$ is a complex function of force.  (c)~Comparison of RFDs for reversible (solid blue line) and irreversible unbinding (dashed red line).  The unbinding forces in~(a) are distributed according to $p[F]$, which is much broader than the distribution $\pirr[F]$ for strictly irreversible unbinding.  }
\label{fig:force_extension_curves}
\end{center}
\end{figure}

The fact that Eq.~\eqref{eq:rate-equations} describes a closed system, which can only transition between the bound and the unbound state, \emph{i.e.}, $B(t) + U(t) = 1$, can be used to decouple the equations, giving
\begin{equation}\label{eq:decoupled-rate-equations}
\begin{aligned}
\dot{B}(t) & = - [ \koff(t) + \kon(t) ] B(t) + \kon(t) \, ,
\\
U(t) & = 1 - B(t) \, .  
\end{aligned}
\end{equation}
The formal solution of Eq.~\eqref{eq:decoupled-rate-equations} reads~\cite{Talkner2003}
\begin{equation}\label{eq:formal-solution}
B(t) = B(t') \e^{-K(t,t')} + \int_{t'}^{t} \ddf{\tau} \e^{-K(t,\tau)} \kon(\tau) \, , 
\end{equation}
where $t \geq t'$ and the function $K$ is defined as
\begin{equation}\label{eq:K-function}
\begin{aligned}
K(t,t') & = \int_{t'}^{t} \ddf{\tau} [ \koff(\tau) + \kon(\tau) ] \, , 
\\
K(t) & \equiv K(t,0) \, , 
\\
K(t,t') & = K(t) - K(t') \, .  
\end{aligned}
\end{equation}
If recrossings are forbidden, the unbinding process becomes irreversible and Eq.~\eqref{eq:formal-solution} reduces to
\begin{equation*}
\Birr(t) = \Birr(t') \exp \bigg( - \int_{t'}^{t} \ddf{\tau} \koff(\tau) \bigg) \, ,
\end{equation*}
This expression can, in combination with the identity
\begin{equation}\label{eq:irreversible-rfd}
\pirr[F] \mathrm{d}F = \koff(t) \Birr(t) \mathrm{d}t \, , 
\end{equation}
be used to calculate the RFD $\pirr[F]$ in closed form for linear force-ramp protocols $F(t) \propto \smash{\dot{F}} t$ and popular rate models, such as the Bell rate~\cite{Bell1978} or the Dudko-Hummer-Szabo (DHS) rate~\cite{DudkoHummer2006}.  

If multiple unbinding and rebinding events occur before the bond dissociates permanently, as is illustrated in the exemplary force-extension curves of Figs.~\ref{fig:force_extension_curves}a and~b, the corresponding RFD $p[F]$ not only describes the statistics of the final unbinding force, but of all unbinding forces.  The properly normalized RFD is given by
\begin{equation}\label{eq:reversible-rfd}
\begin{aligned}
p[F] \mathrm{d}F & = \frac{\koff(t) B(t)}{\avg{\Noff}} \mathrm{d}t \, ,
\\
\avg{\Noff} & = \int_{t'}^{\infty} \ddf{\tau} \koff(\tau) B(\tau) \, ,
\end{aligned}
\end{equation}
where the normalization factor $\avg{\Noff}$ corresponds to the average number of unbinding events and $t=t'$ denotes the initiation time of the force protocol.  Figure~\ref{fig:force_extension_curves}c demonstrates that $p[F]$ is much broader and takes significant values at higher rupture forces than $\pirr[F]$, which is why one has to be careful not to interpret data collected at reversible conditions with a theory developed for irreversible rupture.  A corresponding expression for the distribution of rebinding forces and the average number of rebinding events can be constructed analogously to $p[F]$ and $\avg{\Noff}$.  Higher moments of the number of unbinding events can be calculated using the theory of Refs.~\onlinecite{GopichSzabo2003, GopichSzabo2005}, which, \emph{e.g.}, for the second factorial moment of $\Noff$ gives the following expression:
\begin{equation}\label{eq:second-factorial-moment}
\begin{aligned}
& \avg{\Noff (\Noff - 1)} = 2 \int_{0}^{\infty} \ddf{\tau} \koff(\tau)
\\
& \; \times \int_{t'}^{\tau} \ddf{\tau'} \e^{- K(\tau,\tau')} \kon(\tau') \int_{t'}^{\tau'} \ddf{\tau''} \koff(\tau'') B(\tau'') \, .  
\end{aligned}
\end{equation}
A detailed derivation of the moments and the probabilities of observing exactly $n$ unbinding transitions after a certain time $t$ can be found in Appendix~\ref{app:transition-counts}.  

Unfortunately, the integral term in Eq.~\eqref{eq:formal-solution} cannot be computed analytically for the above-mentioned common rate expressions, which is why quantities like $\avg{\Noff}$ and $\avg{\Noff(\Noff - 1)}$ contain nested integrals that are computationally expensive to evaluate.  It is therefore of great interest to find closed-form approximations to Eq.~\eqref{eq:formal-solution} that remain valid in the parameter range where rebinding events are common.  

In what follows, we shall discuss two approximations that hold in opposing limits, namely for vanishingly low and extremely large loading rates $\dot{F}$.  The latter limit was originally treated in Ref.~\onlinecite{Friddle2016} but, as we shall see in Sec.~\ref{sec:approximations}, turns out to be the less interesting of the two, because rebinding rarely occurs in this regime, if at all.

\section{Approximations}\label{sec:approximations}

At this point it is convenient to introduce the auxiliary population 
\begin{equation}\label{eq:equilibrium-population}
\Bqe(t) = \frac{\kon(t)}{\koff(t) + \kon(t)} \, ,
\end{equation}
where the ``qe'' label refers to ``quasi equilibrium''.  With $\smash{k_{\text{on/off}}^{0} = k_{\text{on/off}}(t \leq t')}$ the spontaneous unbinding and rebinding rates, respectively, Eq.~\eqref{eq:equilibrium-population} reduces to the equilibrium population $\smash{\kon^{0}} / ( \smash{\koff^{0}} + \smash{\kon^{0}} )$ required for detailed balance to hold for $t \leq t'$, where the force protocol has not yet been initiated.  With $\Bqe(t)$, we rewrite Eq.~\eqref{eq:formal-solution} as follows:
\begin{equation}\label{eq:bound-population}
\begin{aligned}
B(t)
& = \Bqe(t) + [B(t') -  \Bqe(t')] \e^{-K(t,t')}
\\
& \mathrel{\phantom{=}} - \int_{t'}^{t} \ddf{\tau} \e^{-K(t,\tau)} \dot{B}_{\text{qe}}(\tau) \, .  
\end{aligned}
\end{equation}
In this form, the integral term denotes nonequilibrium corrections to the quasi-equilibrium population $\Bqe(t)$ with correspondingly modified initial conditions.  Equation~\eqref{eq:bound-population} has, in principle, the same structure as Eq.~\eqref{eq:formal-solution}, but is more convenient to work with when trying to find approximate closed-form expressions for $B(t)$.  Numerical approximations to the integral in Eq.~\eqref{eq:formal-solution} can be used to improve on the quasi-equilibrium approximation where needed.  

\subsection{Instantaneous rebinding approximation}

Starting from Eq.~\eqref{eq:bound-population}, Friddle~\cite{Friddle2016} considered the extreme case of a rebinding rate that causes the bond to instantaneously reform if it is broken at some force $F < F_{1/2}$, but is otherwise negligible.  The threshold force $F_{1/2}$ is chosen as the coexistence force at which the unbinding and rebinding rates are equal, and therefore a solution to the transcendental equation
\begin{equation*}
\koff[F_{1/2}] = \kon[F_{1/2}] \, .  
\end{equation*}
For $t' \to - \infty$, $B(t') = 1$, and a force protocol $F(t) = \dot{F} t$ present at all times $t \in \mathbb{R}$, the second term in Eq.~\eqref{eq:bound-population} vanishes, which can also be achieved with a force protocol initiating at $t=t'$ and the requirement that the system starts in equilibrium, \emph{i.e.}, $\smash{B(t') = \Bqe(t')}$.  Furthermore, in the limit of the extreme rebinding kinetics considered by Friddle, one can approximate $\Bqe(t)$ by a Heaviside unit step function $\Theta$, \emph{i.e.},
\begin{equation}\label{eq:heaviside-step-approximation}
\Bqe(t) \approx B(t') \Theta (t_{1/2} - t) = 
\begin{cases}
B(t') \, , & t < t_{1/2}
\\
0 \, , & \text{otherwise}
\end{cases}
\end{equation}
with $t_{1/2} = F_{1/2} / \dot{F}$.   Note that the prefactor $B(t')$ was not explicitly considered in Ref.~\onlinecite{Friddle2016}, but we include it here to account for the situation that the force protocol has a finite initiation time, as mentioned above.  The approximation behind Eq.~\eqref{eq:heaviside-step-approximation} makes it possible to evaluate the integral in Eq.~\eqref{eq:bound-population}, giving
\begin{equation}\label{eq:Friddle-approximation}
\Binst(t) = B(t') \big[ \Theta (t_{1/2} - t) + \Theta (t - t_{1/2}) \e^{-K(t,t_{1/2})} \big] \, .  
\end{equation}
In this approximation, no unbinding events occur below $F_{1/2}$, with $B(t < t_{1/2}) \equiv B(t')$.  If the rebinding rate $\kon[F]$ can be neglected for $F > F_{1/2}$, then the corresponding RFD can be calculated from Eq.~\eqref{eq:irreversible-rfd}, resulting in
\begin{equation}\label{eq:Friddle-rfd}
\begin{aligned}
\pinst[F]
& = \frac{\koff[F]}{\dot{F}} \Theta [F - F_{1/2}]
\\
& \mathrel{\phantom{\approx}} \times \exp \bigg( - \frac{1}{\dot{F}} \int_{F_{1/2}}^{F} \ddf{f} \koff[f] \bigg) \, ,
\end{aligned}
\end{equation}
where the sole effect of rebinding is contained in the parameter $F_{1/2}$.  Equation~\eqref{eq:Friddle-rfd} can be used to rederive the closed-form expression for the mean rupture force in Ref.~\onlinecite{FriddleNoy2012}.

\subsection{Quasi-equilibrium approximation}

For sufficiently slow force protocols, the nonequilibrium component of Eq.~\eqref{eq:bound-population} becomes negligible and the relative population in the bound state is approximately given by
\begin{equation}\label{eq:equilibrium-approximation}
B(t) \approx \Bqe(t) + [B(t') -  \Bqe(t')] \e^{-K(t,t')} \, .  
\end{equation}
Equation~\eqref{eq:bound-population} coincides with Eq.~\eqref{eq:equilibrium-approximation} whenever $\koff$ and $\kon$ are constant, which is why we shall refer to the latter as the \emph{quasi-equilibrium approximation} in the remainder of the paper.  

As before, choosing either $\smash{B(t') = \Bqe(t')}$ for a force protocol initiated at $t=t'$ or $\smash{B(t') = 1}$ for $t' \to - \infty$ and a monotonically increasing force protocol makes the second term in Eq.~\eqref{eq:equilibrium-approximation} vanish, giving
\begin{equation}\label{eq:equilibrium-approximation-specific}
B[F] \approx \Bqe[F] = \frac{\kon[F]}{\koff[F] + \kon[F]} \, .  
\end{equation}
Note that this expression is independent of the loading rate $\smash{\dot{F}}$ when evaluated using quasistatic rate models, \emph{i.e.}, models whose rates are only force dependent and do not explicitly vary with $\smash{\dot{F}}$, such as the Bell rate or the DHS rate.  The corresponding average number of unbinding events
\begin{equation}\label{eq:equilibrium-unbinding-number}
\avg{\Noff^{\text{qe}}}[\dot{F}] = \int_{0}^{\infty} \ddf{f} \frac{1}{\dot{F}} \frac{\koff[f] \kon[f]}{\koff[f] + \kon[f]}
\end{equation}
scales inversely with the loading rate for linear force protocols, which implies that pulling ten times faster results in ten times fewer unbinding events on average.  The quasi-equilibrium RFD, which is given by
\begin{equation}\label{eq:equilibrium-rfd}
\pqe[F] = \frac{1}{\dot{F} \avg{\Noff^{\text{qe}}}[\dot{F}]} \frac{\koff[F] \kon[F]}{\koff[F] + \kon[F]} \, , 
\end{equation}
is therefore also independent of $\smash{\dot{F}}$.  Note that the prefactor of Eq.~\eqref{eq:equilibrium-rfd} is independent of $F$ and therefore only needs to be computed once when $\pqe[F]$ is evaluated for multiple rupture forces.  

\begin{figure}[t!]
\begin{center}
\includegraphics{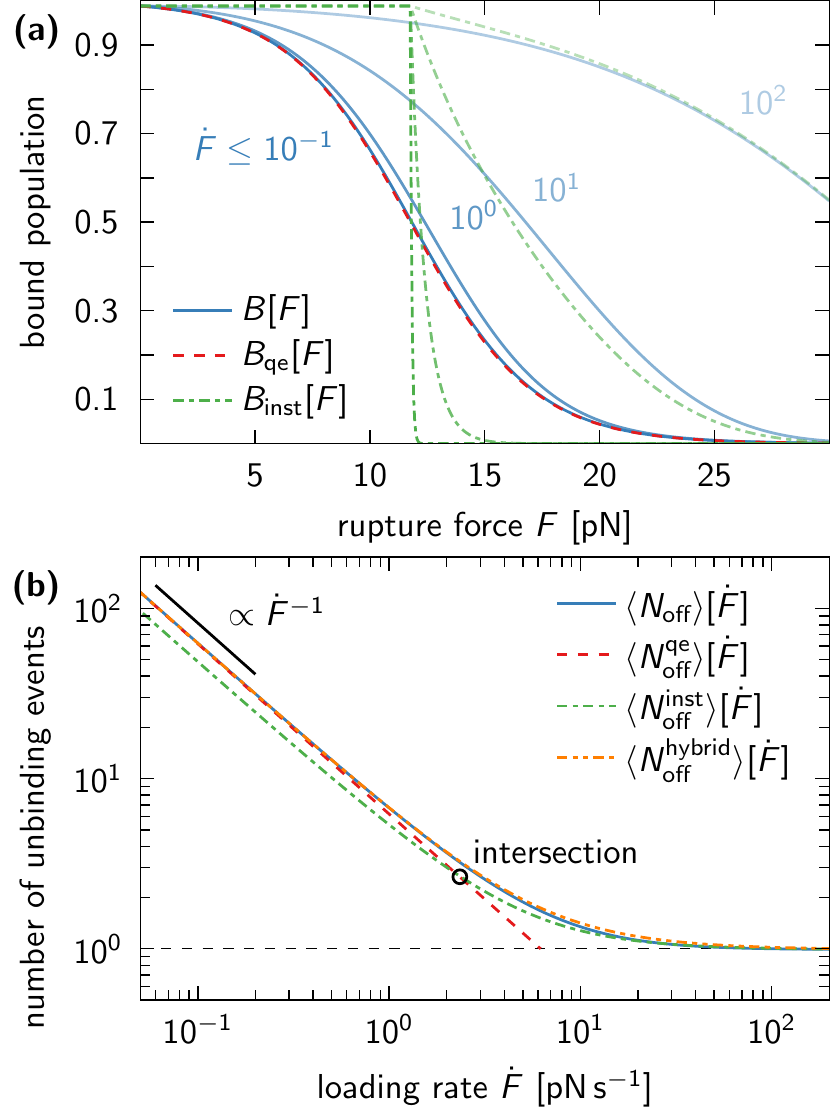}
\caption{Validity range of approximations.  (a)~Relative population in the bound state as a function of the applied force $F = \smash{\dot{F}} t$, plotted for different loading rates $\smash{\dot{F}}$.  The quasi-equilibrium approximation [Eq.~\eqref{eq:equilibrium-approximation-specific}, red dashed line] is independent of $\dot{F}$ and captures well the low-loading-rate behavior of the exact solution [Eq.~\eqref{eq:bound-population}, blue solid lines], whereas the instantaneous rebinding approximation [Eq.~\eqref{eq:Friddle-approximation}, green dashed-dotted lines] only becomes applicable at high loading rates.  (b)~Average number of unbinding events as a function of the loading rate.  The two approximations [Eq.~\eqref{eq:equilibrium-unbinding-number}, red dotted line, and Eq.~\eqref{eq:reversible-rfd} evaluated for $\Binst(t)$, green dashed-dotted line] correctly reproduce the asymptotic limits of $\avg{\Noff}$ (blue solid line), respectively.  Their point of intersection (open black circle) lies just below $\avg{\Noff} = 3$.  The hybrid expression in Eq.~\eqref{eq:hybrid-unbinding-number} (orange dashed-double-dotted line) interpolates between the two limiting regimes and holds for all loading rates.  All equations were evaluated using the Bell-like rates of Eq.~\eqref{eq:Bell-rates} with $\beta^{-1} = \SI{4}{\pico \newton \nano \meter}$, $\dxoff = \SI{0.5}{\nano \meter}$, $\koff^{0} \approx \SI{0.18}{\per \second}$, $\dxon = \SI{1.0}{\nano \meter}$ and $\kon^{0} \approx \SI{14.9}{\per \second}$.  }
\label{fig:friddle_vs_equilibrium}
\end{center}
\end{figure}

Figure~\ref{fig:friddle_vs_equilibrium}(a) explores the validity of the quasi-equilibrium approximation [Eq.~\eqref{eq:equilibrium-approximation-specific}] and Friddle's instantaneous rebinding approximation [Eq.~\eqref{eq:Friddle-approximation}] by comparing them directly to the exact solution [Eq.~\eqref{eq:bound-population} with $\smash{B(t') = \Bqe(t')}$] at different loading rates.  We thereby assume Bell-like binding and unbinding rates, namely
\begin{equation}\label{eq:Bell-rates}
\begin{aligned}
\koff[F] & = \koff^{0} \e^{\beta \dxoff F} \, ,
\\
\kon[F] & = \kon^{0} \e^{- \beta \dxon F} \, ,
\end{aligned}
\end{equation}
and a linear force protocol $F(t) = \dot{F} t$.  Here, $\smash{\Delta x_{\text{off/on}}}$ denote the distances from the bound and unbound state to the transition state, respectively, and $\beta = \smash{(\kB T)^{-1}}$ is the inverse thermal energy scale with Boltzmann constant $\kB$ and absolute temperature $T$.  In the opposing limits of extremely low and high loading rates, the bound population is well described by $\smash{\Bqe[F]}$ and $\smash{\Binst[F]}$, respectively.  The average numbers of unbinding events observed in these limits are very different, as demonstrated in Fig.~\ref{fig:friddle_vs_equilibrium}(b).  While the instantaneous rebinding approximation only seems appropriate for $\avg{\Noff} \approx 1$, the quasi-equilibrium approximation holds for arbitrarily large $\avg{\Noff}$.  The two curves cross at $\avg{\Noff} \approx 2.65$ for the parameter values considered here, meaning that the quasi-equilibrium approximation outperforms the instantaneous rebinding approximation in situations where three or more unbinding events are observed on average.  Note that the curves have been truncated below the value of $\avg{\Noff} = 1$, because we always expect to observe at least one unbinding event.  For loading rates where $\avg{\Noff}$ is predicted to take values less than 1, the simpler theory of irreversible rupture should be used.  

By combining the two approximations, the following expression can be constructed:
\begin{equation}\label{eq:hybrid-unbinding-number}
\begin{aligned}
\avg{\Noff^{\text{hybrid}}} & [\dot{F}] = \int_{0}^{\infty} \ddf{f} \frac{\koff[f] \Bqe[f]}{\dot{F}}
\\
& + \Bqe(t') \int_{F_{1/2}}^{\infty} \ddf{f} \frac{\koff[f] \e^{-K[f,F_{1/2}]}}{\dot{F}} \, ,
\end{aligned}
\end{equation}
which nicely interpolates between the two limits and, unlike Eq.~\eqref{eq:reversible-rfd}, does not require the evaluation of nested integrals.

\subsection{Effects of rebinding on experimental observables at low loading rates}

In practice, poor time resolution might erase certain features from force-extension curves with multiple unbinding-rebinding transitions, like the ones in Figs.~\ref{fig:force_extension_curves}a and~b.  In the extreme case, one might therefore only
resolve a single unbinding event.  However, the effects of rebinding may still affect specific experimental observables, such as the most probable rupture force, which corresponds to the mode $\Fmode$ of the RFD and must satisfy 
\begin{equation}\label{eq:mode-definition}
\frac{\mathrm{d}p[F]}{\mathrm{d} F} \bigg\vert_{\mathrlap{F = \Fmode}} \mathop{=}^{!} 0 \, .  
\end{equation}
Under the quasi-equilibrium approximation, which should be valid in the limit of small loading rates, the RFD is given by Eq.~\eqref{eq:equilibrium-rfd}, for which Eq.~\eqref{eq:mode-definition} reduces to
\begin{equation*}
\frac{\koff'[\Fmode]}{\koff[\Fmode]^{2}} = - \frac{\kon'[\Fmode]}{\kon[\Fmode]^{2}}
\end{equation*}
with $\smash{k_{\text{off/on}}'[F]} = \smash{\mathrm{d}k_{\text{off/on}}[F] / \mathrm{d}F}$.  The equation above can be solved for the Bell-like rates of Eq.~\eqref{eq:Bell-rates}, resulting in the following loading-rate independent value:
\begin{equation}\label{eq:force-spectrum-offset}
\Fmode = F_{1/2} + \frac{\ln (\dxoff / \dxon)}{\beta (\dxoff + \dxon)} \, ,
\end{equation}
which is offset by a constant from the coexistence force $F_{1/2} = \smash{\beta^{-1} \ln(\kon^{0} / \koff^{0}) / (\dxoff + \dxon)}$.  In contrast to the case of irreversible unbinding, where the mode decreases with $\smash{\ln(\dot{F})}$ and vanishes at a finite loading rate value~\cite{EvansRitchie1997}, the mode of $p[F] \approx \pqe[F]$ converges to a fixed value as $\smash{\dot{F} \to 0}$.

The force spectrum therefore does not converge to $F=0$ for $\smash{\dot{F}} \to 0$, as expected for irreversible unbinding, but instead converges towards $F = \Fmode$.  

In Ref.~\onlinecite{FriddleNoy2012} a similar result was derived for the mean rupture force $\smash{\avg{F}[\dot{F}]}$, while indirectly employing the instantaneous rebinding approximation.  The authors of the study found that $\smash{\avg{F}[\dot{F} \to 0]} = F_{1/2}$, which coincides with our estimate [Eq.~\eqref{eq:force-spectrum-offset}] whenever $\dxoff \approx \dxon$, but otherwise can deviate significantly from $\Fmode$.

\section{Extracting rates and transition state locations from pulling experiments}\label{sec:data-fitting}

The data recorded in SMFS experiments are force-extension curves, showing transitions between discrete states.  We assume that each point of the curve can be assigned to one of the two possible states, \emph{i.e.}, either the bound ($s = b$) or the unbound ($s = u$) state.  This gives rise to a time series
\begin{equation}\label{eq:data}
\{ s_{n} \}_{n=0}^{N} = \{ b, b, \dots, b, u, u, \dots, u, b, b, \dots \}
\end{equation}
accompanied by the forces $\smash{\{ F_{n} \}_{n=0}^{N}}$ applied at each time instance.  Such data can be analyzed in various ways, \emph{e.g.}, one can construct a likelihood~\cite{GetfertReimann2007, BullerjahnSturm2014} using the RFD in Eq.~\eqref{eq:equilibrium-rfd}, or analyze the dwell times in each state~\cite{Talkner2003}.  However, these approaches involve integrals over the rates, which have to be evaluated numerically for more complex rate expressions than the Bell rate or the DHS rate, just to name a few, and the nonlinear force protocols commonly realized in experiments (see Fig.~\ref{fig:force_extension_curves}b).  This increases the numerical cost of data fitting, because said integrals have to be re-evaluated each time the parameters are varied.  

For this reason, we instead opt here for an alternative approach, where we exploit the ``stroboscopic'' nature of the data set and assume that the time step between subsequent observations is sufficiently small to facilitate a short-time expansion of difficult-to-evaluate functions and integrals.  This gives rise to a set of maximum likelihood estimators that are asymptotically exact for small time steps.

\subsection{Maximum likelihood estimators}\label{subsec:maximum-likelihood}

The data structure in Eq.~\eqref{eq:data}, combined with the Markov-assumption that all state transitions (even transitions into the current state) are history-independent,  allows us to decompose the joint probability of observing the time series above in terms of the state-transition probabilities as follows~\cite{Hummer2005}:
\begin{equation}\label{eq:conditional-probability}
\Pr (\{ s_{n} \}) = \Pr(s_{0}, t_{0}) \prod_{n=1}^{N} \Pr (s_{n}, t_{n} \mid s_{n-1}, t_{n-1} ) \, .  
\end{equation}
Here, we only consider trajectories that start in the bound state, \emph{i.e.}, $\Pr(s_{0} = b, t_{0}) = 1$.  The trajectories are sampled at discrete times $t_{n} = n \Delta t$ with a constant time step $\Delta t$ between two subsequent measurements.  $\Pr(s,t \mid s',t')$ denotes the conditional probability of finding the system in state $s$ at time $t$ if it was previously observed in state $s'$ at time $t'$.  

Equation~\eqref{eq:conditional-probability} can be reinterpreted as a likelihood, where the individual components are approximately given by (see Appendix~\ref{app:mle-approach})
\begin{equation}\label{eq:transition-probabilities}
\begin{aligned}
\Pr(b, t_{n} \mid b, t_{n-1}) & \approx 1 - \koff(t_{n-1}) \Delta t \, , 
\\
\Pr(b, t_{n} \mid u, t_{n-1}) & \approx \kon(t_{n-1}) \Delta t \, , 
\\
\Pr(u, t_{n} \mid b, t_{n-1}) & \approx \koff(t_{n-1}) \Delta t \, , 
\\
\Pr(u, t_{n} \mid u, t_{n-1}) & \approx 1 - \kon(t_{n-1}) \Delta t \, , 
\end{aligned}
\end{equation}
for a sufficiently small time step $\Delta t$.  Note that Eq.~\eqref{eq:transition-probabilities} only contains the rates themselves and not their integrals, which circumvents the issues plaguing the analyses of dwell times and rupture forces.  Also note that Eq.~\eqref{eq:transition-probabilities} holds irrespective of the applied loading rate, because it does not depend on the quasi-equilibrium approximation.  The time step $\Delta t$ only needs to be sufficiently small to suppress systematic biases (see further Sec.~\ref{sec:data-analysis}).  

If the rates are modeled Bell-like [Eq.~\eqref{eq:Bell-rates}], then the four-dimensional optimization problem of maximizing the likelihood with respect to the parameters $\{ \dxoff, \koff^{0}, \dxon, \kon^{0} \}$ reduces effectively to two independent one-dimensional problems (see Appendix~\ref{app:mle-approach}), where the negative log-likelihoods
\begin{align}
& \begin{aligned}\label{eq:dxb-eq}
\mathcal{L}(\dxoff \mid \{ s_{n} \}) & = \Nbu \ln \Bigg( \sum_{n=1}^{\Nbb} \e^{\beta \dxoff \Fbb_{n}} \Bigg)
\\
& \mathrel{\phantom{=}} - \sum_{n=1}^{\Nbu} \beta \dxoff \Fbu_{n} \, ,
\end{aligned}
\\
& \begin{aligned}\label{eq:dxu-eq}
\mathcal{L}(\dxon \mid \{ s_{n} \}) & = \Nub \ln \Bigg( \sum_{n=1}^{\Nuu} \e^{- \beta \dxon \Fuu_{n}} \Bigg)
\\
& \mathrel{\phantom{=}} + \sum_{n=1}^{\Nub} \beta \dxon \Fub_{n} \, ,
\end{aligned}
\end{align}
have to be minimized with respect to $\dxoff$ and $\dxon$, respectively, to find the maximum likelihood estimates (MLEs) of these parameters.  Here, $F_{n}^{i \to j}$ denotes the force measured during the $n$-th transition ($n = 1, 2, \dots, N_{i \to j}$) from state $i$ to $j$ and
\begin{equation*}
N = \Nbb + \Nbu + \Nub + \Nuu
\end{equation*}
is the total number of transition counts.  Note that $\Nbu = \Noff$ corresponds to the number of unbinding events.  The remaining two parameters, $\koff^{0}$ and $\kon^{0}$, can be estimated as follows:
\begin{align}
\label{eq:kbu0-eq}
\koff^{0} & = \frac{\Nbu}{\Delta t} \Bigg[ \sum_{n=1}^{\Nbb} \e^{\beta \dxoff \Fbb_{n}} \Bigg]^{-1} \, ,
\\
\label{eq:kub0-eq}
\kon^{0} & = \frac{\Nub}{\Delta t} \Bigg[ \sum_{n=1}^{\Nuu} \e^{- \beta \dxon \Fuu_{n}} \Bigg]^{-1} \, .  
\end{align}
The numerical minimization of Eqs.~\eqref{eq:dxb-eq} and~\eqref{eq:dxu-eq} can be conducted efficiently using robust derivative-free algorithms.  Our own code~\cite{JuliaScript} relies on an implementation of Brent’s method~\cite{Brent1973} provided by the \texttt{Optim} package~\cite{MogensenRiseth2018}.  

\begin{figure*}[t!]
\begin{center}
\includegraphics{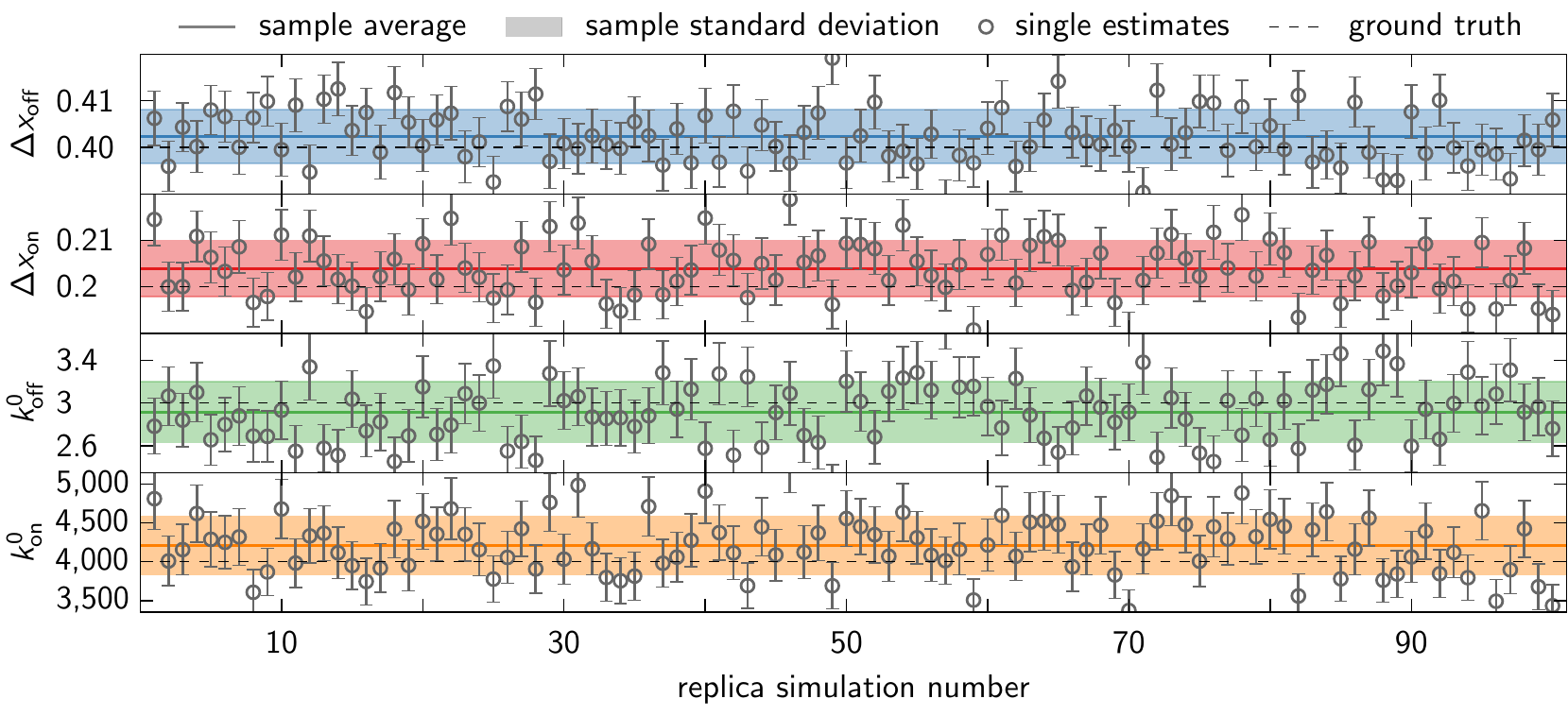}
\caption{Single replica MLEs of linear force-extension curves and their sample statistics.  Each replica simulation ($\smash{\dot{F}} = \SI{100}{\pico \newton \per \second}$, other simulation parameters as specified in Appendix~\ref{app:Gillespie-algorithm}) generated a data set of $M=10$ time series, which was analyzed using the estimators in Eqs.~\eqref{eq:dxb-eq}--\eqref{eq:kub0-eq} at a fixed time step $\Delta t = \SI{0.1}{\micro \second}$.  The individual parameter estimates (gray open circles) scatter around their respective sample average (solid lines), resulting in a sample standard deviation (shaded regions) of comparable size as the corresponding standard error estimate [Eq.~\eqref{eq:mle-errors} with $\delta F^{2} \equiv 0$, error bars].  Slight deviations between sample averages and ground truth values of the parameters (dashed lines) are due to the use of a finite time step, and therefore vanish as $\Delta t \to 0$ (see Fig.~\ref{fig:stroboscopic_estimator_bias}).  }
\label{fig:mle-demonstration}
\end{center}
\end{figure*}

The uncertainties of our parameter estimates can be assessed using the standard errors $\delta \theta_{i} = \sqrt{\var(\theta_{i})}$, which are bounded from below by the Cram\'{e}r-Rao bounds as follows:
\begin{equation*}
\var (\theta_{i}) \geq I_{ii}^{-1} \, .  
\end{equation*}
Here, $I_{ii}^{-1}$ denotes the $i$-th diagonal element of the inverse of the Fisher information matrix $\smash{\mathbf{I}(\vec{\theta})}$, whose elements are given by
\begin{equation*}
I_{ij} = \bigg\langle \frac{\partial^{2} \mathcal{L}(\vec{\theta} \mid \vec{s})}{\partial \theta_{i} \theta_{j}} \bigg\rangle \, .  
\end{equation*}
For the Bell-like rates of Eq.~\eqref{eq:Bell-rates}, we have $\smash{\vec{\theta} = (\dxoff, \koff^{0}, \dxon, \kon^{0})^{T}}$ and $\smash{\mathbf{I}(\vec{\theta})}$ is therefore a $4 \times 4$ matrix.  However, within the low-order approximation of the propagators in Eq.~\eqref{eq:transition-probabilities} the parameters of the binding rate are assumed to be independent of the parameters of the unbinding rate.  The Fisher information matrix then decomposes into two independent $2 \times 2$ matrices, namely $\mathbf{I}_{\text{off}}(\dxoff, \koff^{0})$ and $\mathbf{I}_{\text{on}}(\dxon, \kon^{0})$ with elements (see further Appendix~\ref{app:fim})
\begin{align*}
I_{11}^{\text{off}} & = \beta^{2} \koff^{0} \Delta t \, \e^{(\beta \dxoff \delta F)^{2} / 2} \sum_{n=1}^{\Nbb} \e^{\beta \dxoff \Fbb_{n}}
\\
& \mathrel{\phantom{=}} \times \big[ \delta F^{2} + (\Fbb_{n} + \beta \dxoff \delta F^{2})^{2} \big] \, ,
\\
I_{12}^{\text{off}} & = \beta \Delta t \, \e^{(\beta \dxoff \delta F)^{2} / 2} \sum_{n=1}^{\Nbb} \e^{\beta \dxoff \Fbb_{n}}
\\
& \mathrel{\phantom{=}} \times (\Fbb_{n} + \beta \dxoff \delta F^{2}) \, ,
\\
I_{22}^{\text{off}} & = \frac{\Nbu}{(\koff^{0})^{2}} \, ,
\\
I_{11}^{\text{on}} & = \beta^{2} \kon^{0} \Delta t \, \e^{(\beta \dxon \delta F)^{2} / 2} \sum_{n=1}^{\Nuu} \e^{- \beta \dxon \Fuu_{n}}
\\
& \mathrel{\phantom{=}} \times \big[ \delta F^{2} + (\Fuu_{n} - \beta \dxon \delta F^{2})^{2} \big] \, ,
\\
I_{12}^{\text{on}} & = - \beta \Delta t \, \e^{(\beta \dxon \delta F)^{2} / 2} \sum_{n=1}^{\Nuu} \e^{- \beta \dxon \Fuu_{n}}
\\
& \mathrel{\phantom{=}} \times (\Fuu_{n} - \beta \dxon \delta F^{2}) \, ,
\\
I_{22}^{\text{on}} & = \frac{\Nub}{(\kon^{0})^{2}} \, .  
\end{align*}
Here, $\delta F^{2}$ denotes the experimentally determined mean-squared fluctuations of the applied force, which we assume to be constant.  The elements of the reduced information matrices can be used to estimate the standard errors $\delta \theta_{i} = \sqrt{\var(\theta_{i})}$ from below, giving
\begin{equation}\label{eq:mle-errors}
\begin{aligned}
\delta \dxoff & \geq \sqrt{\frac{I_{22}^{\text{off}}}{I_{11}^{\text{off}} I_{22}^{\text{off}} - (I_{12}^{\text{off}})^{2}}} \, ,
\\
\delta \koff^{0} & \geq \sqrt{\frac{I_{11}^{\text{off}}}{I_{11}^{\text{off}} I_{22}^{\text{off}} - (I_{12}^{\text{off}})^{2}}} \, ,
\\
\delta \dxon & \geq \sqrt{\frac{I_{22}^{\text{on}}}{I_{1,1}^{\text{on}} I_{22}^{\text{on}} - (I_{12}^{\text{on}})^{2}}} \, ,
\\
\delta \kon^{0} & \geq \sqrt{\frac{I_{11}^{\text{on}}}{I_{11}^{\text{on}} I_{22}^{\text{on}} - (I_{12}^{\text{on}})^{2}}} \, ,
\end{aligned}
\end{equation}
where equality can be assumed for large data sets containing multiple transition events.

\begin{figure*}[t!]
\begin{center}
\includegraphics{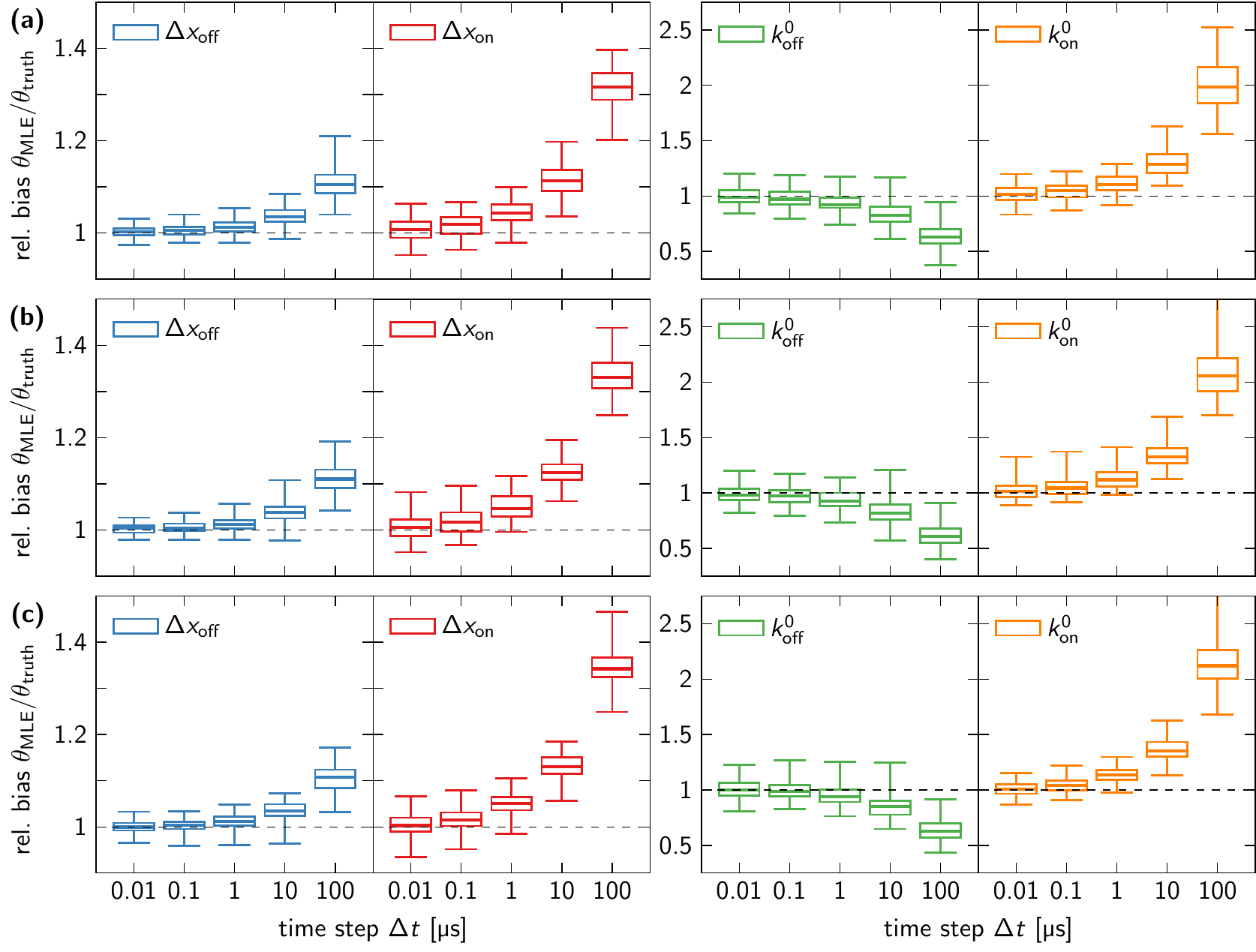}
\caption{Estimating relative bias in the rate parameter MLEs with respect to the loading rate $\smash{\dot{F}}$ and time step $\Delta t$.  To imitate the experimental situation of limited instrumental time resolution, we lengthen the time step between subsequent observations in our data sets and analyze the resulting time series using Eqs.~\eqref{eq:dxb-eq}--\eqref{eq:kub0-eq}.  Generally, the accuracy of the parameter estimates decreases with increasing $\Delta t$.  We observe this trend at three different loading rates, (a)~$\smash{\dot{F}} = 10$, (b)~$\smash{\dot{F}} = 100$ and (c)~$\smash{\dot{F}} = \SI{1000}{\pico \newton \per \second}$, in exactly the same fashion, thus indicating that the effect is $\smash{\dot{F}}$-independent.  Note that we did not exclude any outliers, so the whiskers of the box plots mark the maximum and minimum values of each sample.  }
\label{fig:stroboscopic_estimator_bias}
\end{center}
\end{figure*}

\subsection{Analysis of synthetic force-extension curves}\label{sec:data-analysis}

To demonstrate the applicability of our results, we used them to analyze synthetic force-extension curves, generated by a Gillespie stochastic simulation algorithm for time-dependent rates~\cite{ThanhPriami2015} (see Appendix~\ref{app:Gillespie-algorithm} for details).  For convenience, we only considered force protocols with $\smash{\dot{F}} = \const$, but the estimators in Eqs.~\eqref{eq:dxb-eq}--\eqref{eq:kub0-eq} are applicable to arbitrary monotonically increasing $F(t)$.  To accommodate for the fact that more unbinding-rebinding transitions are observed at slower loading rates prior to permanent dissociation, we varied the number of pulling experiments $M$ to keep the total number of average unbinding events $M \times \avg{\Noff}$ approximately constant for each data set.  Our combinations of $\smash{\dot{F}}$ and $M$ are listed in Table~\ref{tab:simulation-parameters}, along with the number of replica simulations generated to calculate sample statistics.  All simulations were conducted using the ``ground truth'' parameter values
\begin{equation}\label{eq:ground-truth}
\begin{aligned}
\dxoff^{\text{truth}} & = \SI{0.4}{\nano \meter} \, , & \dxon^{\text{truth}} & = \SI{0.2}{\nano \meter} \, ,
\\
\koff^{0,\text{truth}} & = \SI{3}{\per \second} \, , & \kon^{0,\text{truth}} & = \SI{4000}{\per \second} \, .  
\end{aligned}
\end{equation}

\begin{table}[h]
\caption{Replica simulation specifications.  With increasing loading rate $\smash{\dot{F}}$, the (sample) average number of unbinding events $\overline{N}_{b \to u} \approx \avg{\Noff}$ observed in each pulling experiment decreased, so the number of experiments $M$ in each replica had to be adjusted to compensate.  A total of $N_{\text{repl}}$ replica simulations were run for each considered loading rate.}
\begin{tabularx}{\linewidth}{X X X X r}
$\smash{\dot{F}}~[\si{\pico \newton \per \second}]$ & $\overline{N}_{b \to u}$ & $M$ & $M \times \overline{N}_{b \to u}$ & $N_{\text{repl}}$
\\
\toprule
10 & 1069.9 & 1 & 1070 & 100 \\
100 & 107.40 & 10 & 1074 & 100 \\
1000 & 11.28 & 100 & 1128 & 100
\\
\bottomrule
\end{tabularx}
\label{tab:simulation-parameters}
\end{table}

The time series of each replica data set were analyzed separately, using Eqs.~\eqref{eq:dxb-eq}--\eqref{eq:kub0-eq} to estimate the model parameters $\{ \dxoff, \koff^{0}, \dxon, \kon^{0} \}$, and Eq.~\eqref{eq:mle-errors} to estimate the corresponding uncertainties $\{ \delta\dxoff, \delta\koff^{0}, \delta\dxon, \delta\kon^{0} \}$.  Note that the applied force did not fluctuate in our simulations, so $\delta F^{2} \equiv 0$.  The sample average and sample standard deviation of the replica MLEs were calculated as follows:
\begin{equation*}
\begin{aligned}
\overline{\theta} & = \frac{1}{N_{\text{repl}}} \sum_{n=1}^{N_{\text{repl}}} \theta_{n} \, , 
\\
\delta \overline{\theta} & = \sqrt{ \sum_{n=1}^{N_{\text{repl}}} \frac{(\theta_{n} - \overline{\theta})^{2}}{N_{\text{repl}} - 1} } \, , 
\end{aligned}
\end{equation*}
for $\theta \in \{ \dxoff, \koff^{0}, \dxon, \kon^{0} \}$, and are plotted against the single replica MLEs and associated uncertainties of the $\smash{\dot{F}} = \SI{100}{\pico \newton \per \second}$ simulations in Fig.~\ref{fig:mle-demonstration}.  As expected, the standard error predictions for the replica MLEs mostly coincide with the sample standard deviation, which demonstrates the usefulness of Eq.~\eqref{eq:mle-errors} to gauge the uncertainty of parameter estimates, especially when only a single or few time series are available for the analysis.  

\begin{figure}[t!]
\begin{center}
\includegraphics{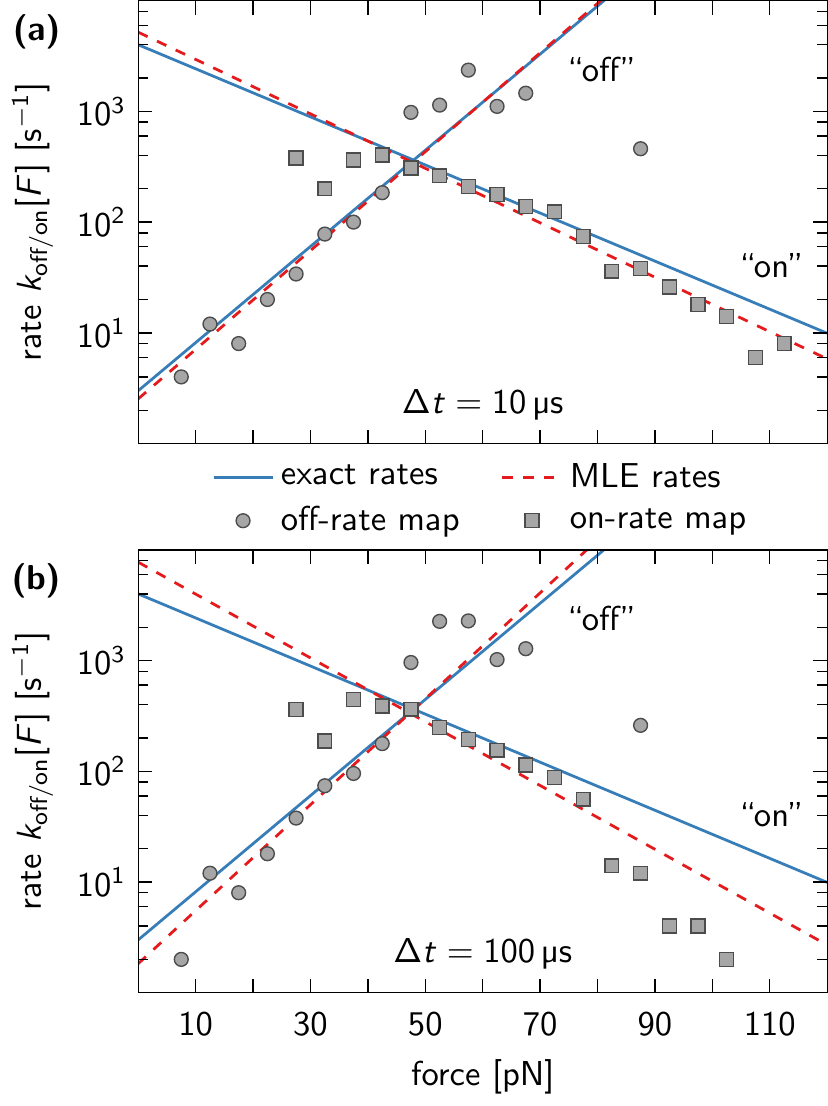}
\caption{Comparing rate map predictions to MLEs at limited response times.  A single set of force-extension curves ($\smash{\dot{F}} = \SI{100}{\pico \newton \per \second}$, $M = 10$) was analyzed, on the one hand, using the rate-map method of Ref.~\onlinecite{ZhangDudko2013} (gray circles and squares) and, on the other hand, via the estimators in Eqs.~\eqref{eq:dxb-eq}--\eqref{eq:kub0-eq} (red dashed lines).  The MLEs are as follows: (a)~$\dxoff = \SI{0.411 \pm 0.007}{\nano \meter}$, $\koff^{0} = \SI{2.5 \pm 0.3}{\per \second}$, $\dxon = \SI{0.227 \pm 0.006}{\nano \meter}$, $\kon^{0} = \SI{5200 \pm 500}{\per \second}$ for $\Delta t = \SI{10}{\micro \second}$, and (b)~$\dxoff = \SI{0.441 \pm 0.009}{\nano \meter}$, $\koff^{0} = \SI{1.8 \pm 0.2}{\per \second}$, $\dxon = \SI{0.265 \pm 0.007}{\nano \meter}$, $\kon^{0} = \SI{7700 \pm 800}{\per \second}$ for $\Delta t = \SI{100}{\micro \second}$.  Deviations from the exact rates (blue solid lines) increase as the time resolution gets poorer.  }
\label{fig:rate_map}
\end{center}
\end{figure}

We also investigated the influence of limited instrumental response times on our MLEs.  Our estimators rest on a short-time approximation of the transition probabilities [Eq.~\eqref{eq:transition-probabilities}], which can introduce biases for long observation intervals $\Delta t$~\cite{JacobsonPerkins2020}.  We therefore increased the time step $\Delta t$ between observations to generate time series with a limited time resolution that we then subsequently analyzed.  In Fig.~\ref{fig:stroboscopic_estimator_bias} we demonstrate how the distributions of the MLEs vary with $\Delta t$ at different loading rates.  At $\Delta t = \SI{0.01}{\micro \second}$, the estimators are virtually unbiased, whereas at larger response times the relative bias can result in estimates up to twice as large as the ground-truth values.  This behavior is also observed in the corresponding rate maps, as seen in Fig.~\ref{fig:rate_map}, where we compare the binned rate estimates obtained by the method proposed in Ref.~\onlinecite{ZhangDudko2013} to Eq.~\eqref{eq:Bell-rates}, evaluated for the ground-truth values [Eq.~\eqref{eq:ground-truth}] and the associated MLEs, respectively.  The two methods are fairly consistent, even for large time steps, because they both rely on the fact that the probability to observe an unbinding or rebinding event within the time interval $[t, t+\mathrm{d}t]$ corresponds to either $\koff(t) \mathrm{d}t$ or $\kon(t) \mathrm{d}t$.  They also complement each other in the sense that rate maps give nonparametric predictions for the force-dependence of the rates, whereas our theory provides an optimal and straightforward way to fit explicit rate expressions to the data.

\section{Conclusions}\label{sec:conclusions}

We have explored and quantified the effects of rebinding kinetics on single-molecule force spectroscopy experiments under external loads, particularly in the near-equilibrium limit of slow pulling, where multiple unbinding-rebinding transitions can be observed prior to a permanent dissociation of the bond.  In this limit, the system can be assumed to be close to equilibrium, which resolves the inherent problem of nonequilibrium ramping typically associated with driven force protocols.  Furthermore, the quasi-equilibrium approximation allowed us to derive a closed-form expression for the relative population in the bound state [Eq.~\eqref{eq:equilibrium-approximation} for arbitrary initial conditions, Eq.~\eqref{eq:equilibrium-approximation-specific} for equilibrium initial conditions], and a near closed-form expression for the corresponding rupture force distribution [Eq.~\eqref{eq:equilibrium-rfd}].  

As one of the limitations, our approach assumes an explicit dependence of the applied force on time, and thus ignores fluctuations both in the measurement apparatus and in the attached molecular construct~\cite{HummerSzabo2003}, as well as finite relaxation times of the apparatus and linker~\cite{CossioHummer2015}.  Including such effects will require more elaborate theoretical formulations, \emph{e.g.}, by going from the rate equations of Eq.~\eqref{eq:rate-equations} to low-dimensional diffusion models~\cite{HummerSzabo2003, CossioHummer2015}.  Another limitation is the low-order approximation of the propagators in Eq.~\eqref{eq:transition-probabilities}.  This approximation can be improved upon by discrete integration of the kinetics between the sampled state observations with a short time step, $\delta t < \Delta t$.  However, this finer time integration or the use of diffusion models will result in more complex likelihood functions that are less amenable to analytical treatment and numerical optimization.  Finally, on the experimental side, one may not be able to deduce the state at the sampled times $t_{n}$ with confidence.  If one uses probabilities to express the confidence in the respective state observations, one again ends up with a modified likelihood function less amenable to optimization.  Therefore, we concentrated here on the unbinding-rebinding kinetics in its simplest form.  

We demonstrated how a short-time expansion of a state propagator, whose kinetics is described by Eq.~\eqref{eq:rate-equations}, can be used to analyze experimental force-extension curves with multiple unbinding-rebinding transitions via the principle of maximum likelihood.  The resulting likelihood can be evaluated for rates with arbitrarily complex functional forms, and does not explicitly depend on the applied force protocol.  In the special case of Bell-like rates, the four-dimensional optimization problem of maximizing the likelihood with respect to the model parameters reduces to two separate one-dimensional ones.  This gives rise to a numerically efficient scheme to estimate the parameters and their associated uncertainties, which we have implemented in an open-source data analysis package~\cite{JuliaScript}.  We have validated our approach against extensive kinetic simulation data, which also revealed potential biases in the analysis of experiments with limited time resolution.  In cases where the small bias resulting from the use of short-time approximated state propagators in the likelihood is not acceptable, one can use the exact propagators [Eq.~\eqref{eq:formal-solution}] obtained by solving the two-state kinetic model of Eq.~\eqref{eq:rate-equations}. For rate models where analytical calculations are not feasible, the state propagators can be calculated numerically using standard numerical solvers for ordinary differential equations.  Our maximum-likelihood estimators for unbinding and rebinding rates complement existing methods, in particular rate maps~\cite{ZhangDudko2013}, and should find use in a range of applications, from single-molecule force spectroscopy~\cite{Noy2011, YuJacobson2020, JacobsonPerkins2020} to the nanoscopic chemical imaging of surfaces using atomic force microscopes~\cite{MullerDumitru2021}.

\begin{acknowledgments}

We thank Dr.~Attila Szabo for countless insightful and stimulating discussions.  This research was supported by the Max Planck Society and the LOEWE (Landes-Offensive zur Entwicklung Wissenschaftlich-ökonomischer Exzellenz) program of the state of Hesse conducted within the framework of the MegaSyn Research Cluster.  

\end{acknowledgments}

\begin{appendix}

\section{Statistics of transition counts}\label{app:transition-counts}

Let us consider a two-state system, whose kinetics is described via Eq.~\eqref{eq:rate-equations}.  We are interested in the statistics of the number of unbinding events, as the system evolves from an initial bound state at time $t=0$ to some uncertain state at time $t$.  This problem can be tackled with the theory introduced in Refs.~\onlinecite{GopichSzabo2003, GopichSzabo2005}, which we adapt to our problem with time-dependent rates.  

As a first step, we construct the generating function of the distribution of unbinding transition counts.  One thereby considers a modification of the rate matrix behind Eq.~\eqref{eq:rate-equations}, namely
\begin{equation*}
\mathbf{K}(t \mid \lambda) = 
\begin{pmatrix}
- \koff(t) & \kon(t)
\\
\lambda \koff(t) & - \kon(t)
\end{pmatrix} \, , 
\end{equation*}
where $\lambda$ plays the role of a so-called counting parameter.  This name originates from the fact that a system, which evolves according to $\mathbf{K}(t \mid \lambda)$, picks up a factor $\lambda$ every time an unbinding transition occurs.  The corresponding rate equations,
\begin{equation}\label{eq:lambda-rate-equations}
\frac{\mathrm{d}}{\mathrm{d}t}
\begin{pmatrix}
P(t) \\ Q(t)
\end{pmatrix} = \mathbf{K}(t \mid \lambda)
\begin{pmatrix}
P(t) \\ Q(t)
\end{pmatrix} \, ,
\end{equation}
can be formally solved to give
\begin{equation*}
\begin{pmatrix}
P(t) \\ Q(t)
\end{pmatrix} = \mathcal{T} \Bigg\{ \exp \bigg( - \int_{0}^{t} \ddf{\tau} \mathbf{K}(\tau \mid \lambda) \bigg) \Bigg\} \begin{pmatrix}
P(0) \\ Q(0)
\end{pmatrix} \, .  
\end{equation*}
Here, the time-ordering operator $\mathcal{T}$ gives rise to a path-ordered exponential, which is applied to the vector of initial states.  According to Gopich and Szabo~\cite{GopichSzabo2003, GopichSzabo2005}, the sought-after generating function is defined as the summed up probabilities, \emph{i.e.},
\begin{equation}\label{eq:generating-function-1}
G(\lambda, t) = 
\begin{pmatrix}
1 \\ 1
\end{pmatrix}^{T}
\begin{pmatrix}
P(t) \\ Q(t)
\end{pmatrix} = P(t) + Q(t) \, ,
\end{equation}
and satisfies the key relation
\begin{equation*}
G(\lambda, t) = \sum_{n = 0}^{\infty} \Pr(n \mid t) \lambda^{n} \, ,
\end{equation*}
where the coefficients $\Pr(n \mid t)$ give the probabilities of observing exactly $n$ unbinding transitions after a certain time $t$.  

Next, we use a perturbation expansion in $\lambda = 1 + \varepsilon$ with $\varepsilon \to 0$ to solve Eq.~\eqref{eq:lambda-rate-equations}, namely
\begin{equation}\label{eq:perturbation-ansatz}
\begin{aligned}
P(t) & = P_{0}(t) + \varepsilon P_{1}(t) + \varepsilon^{2} P_{2}(t) + \dots \, , 
\\
Q(t) & = Q_{0}(t) + \varepsilon Q_{1}(t) + \varepsilon^{2} Q_{2}(t) + \dots \, , 
\end{aligned}
\end{equation}
where $P_{0}(t) = B(t)$ and $Q_{0}(t) = U(t)$ satisfy Eq.~\eqref{eq:rate-equations}.  By sorting the terms according to powers of $\varepsilon$, we arrive at
\begin{align*}
\dot{P}_{n}(t) & = - \koff(t) P_{n}(t) + \koff(t) Q_{n}(t) \, ,
\\
\dot{Q}_{n}(t) & = \koff(t) P_{n}(t) - \koff(t) Q_{n}(t) + \koff(t) P_{n-1}(t) \, .  
\end{align*}
Using the initial conditions $B(0) = P_{0}(0) = 1$ and $U(0) = Q_{0}(0) = P_{n}(0) = Q_{n}(0) = 0$ for $n \geq 1$, we obtain the following solutions:
\begin{align*}
P_{n}(t) & = \int_{0}^{t} \ddf{\tau} \e^{-K(t,\tau)} \kon(\tau) \int_{0}^{\tau} \ddf{\tau'} \koff(\tau') P_{n-1}(\tau') \, ,
\\
Q_{n}(t) & = - P_{n}(t) + \int_{0}^{t} \ddf{\tau} \koff(\tau) P_{n-1}(\tau) \, ,
\end{align*}
for $n \geq 1$ and Eq.~\eqref{eq:formal-solution} with $t'=0$ for $n=0$.  See Eq.~\eqref{eq:K-function} in the main text for a definition of $K$.  This motivates us to define a function
\begin{align*}
f_{0}(t) & = P_{0}(t) + Q_{0}(t) = 1 \, ,
\\
f_{n}(t) & = P_{n}(t) + Q_{n}(t) = \int_{0}^{t} \ddf{\tau} \koff(\tau) P_{n-1}(\tau) \, , 
\end{align*}
which, according to Eqs.~\eqref{eq:generating-function-1} and~\eqref{eq:perturbation-ansatz}, must satisfy
\begin{equation*}
G(\lambda, t) = P(t) + Q(t) = \sum_{n=0}^{\infty} f_{n}(t) \varepsilon^{n}
\end{equation*}
or, equivalently,
\begin{equation*}
\sum_{n=0}^{\infty} f_{n}(t) \varepsilon^{n} = G(1 + \varepsilon, t) = \sum_{n = 0}^{\infty} \Pr(n \mid t) (1 + \varepsilon)^{n} \, .  
\end{equation*}
It thus becomes apparent that the factorial moments of $n$ after time $t$ can be computed with the help of $f_{n}(t)$ as
\begin{equation}\label{eq:factorial-moments}
\begin{aligned}
\avg{n(n-1) \dots (n-m+1)} & = \frac{\mathrm{d}^{m} G(1+\varepsilon \mid t)}{\mathrm{d} \varepsilon^{m}} \bigg\vert_{\varepsilon = 0}
\\
& = m! f_{n}(t) \, .  
\end{aligned}
\end{equation}
The factorial moments of $\Noff$ can be calculated via Eq.~\eqref{eq:factorial-moments} in the limit of $t \to \infty$.  Its first two moments, characterized by $m=1$ and $m=2$, are given by Eqs.~\eqref{eq:reversible-rfd} and~\eqref{eq:second-factorial-moment} in the main text.  

In cases where Eq.~\eqref{eq:factorial-moments} is not analytically tractable, the factorial moments can be calculated efficiently by applying standard numerical solvers for ordinary differential equations (ODE) to the following system of first-order linear equations:
\begin{align*}
\dot{P}_{n}(t) & = - [\koff(t) + \kon(t)] P_{n}(t) + \kon(t) f_{n}(t) \, ,
\\
\dot{f}_{n+1}(t) & = \koff(t) P_{n}(t) \, ,
\end{align*}
using the initial conditions $P_{0}(0) = 1$ and $f_{n}(0) = P_{n}(0) = 0$, as well as the fact that $f_{0}(t) = 1$ by definition.  

Finally, we want to point out that the probabilities $\Pr(n \mid t)$ of the unbinding transition counts can be determined by quadrature.  For this, we unroll the two-state system onto a line, \emph{i.e.},
\begin{equation*}
0 \mathop{\to}^{k_{0}(t)} 1 \mathop{\to}^{k_{1}(t)} 2 \mathop{\to}^{k_{2}(t)} 3 \mathop{\to}^{k_{3}(t)} \dots \, ,
\end{equation*}
with $k_{n}(t) = \koff(t)$ for $n$ even and $k_{n}(t) = \kon(t)$ for $n$ odd.  The populations $R_{n}$ of this ``unrolled'' process satisfy
\begin{align*}
R_{0}(t) & = \exp \bigg( - \int_{0}^{t} \ddf{\tau} \koff(\tau) \bigg) \, ,
\\
R_{n}(t) & = \int_{0}^{t} \ddf{\tau} \exp \bigg( - \int_{\tau}^{t} \ddf{\tau'} k_{n}(\tau') \bigg) k_{n-1}(\tau) R_{n-1}(\tau) \, ,
\end{align*}
which can be solved recursively for $n \geq 1$.  The population after exactly $n$ unbinding events is then
\begin{equation*}
\Pr(n \mid t) = R_{2n-1}(t) \, .  
\end{equation*}
Again, if the integrals cannot be evaluated analytically, one can use a standard ODE solver for efficient numerical solution.

\section{Derivation of the maximum likelihood estimators}\label{app:mle-approach}

As discussed in Sec.~\ref{subsec:maximum-likelihood}, the idea of the maximum likelihood principle is to reinterpret Eq.~\eqref{eq:conditional-probability} as a likelihood and maximize it with respect to the model parameters.  We thereby need explicit expressions for the conditional probabilities $\Pr(s_{n},t_{n} \mid s_{n-1},t_{n-1})$, which can be constructed from the relative populations $B(t)$ and $U(t) = 1 - B(t)$, giving
\begin{align*}
\Pr( b, t_{n} \mid b, t_{n-1}) & = B(t_{n}) \Big\vert_{\smash{B(t_{n-1}) = 1}} \, , 
\\
\Pr( b, t_{n} \mid u, t_{n-1}) & = B(t_{n}) \Big\vert_{\smash{B(t_{n-1}) = 0}} \, , 
\\
\Pr( u, t_{n} \mid b, t_{n-1}) & = 1 - B(t_{n}) \Big\vert_{\smash{B(t_{n-1}) = 1}} \, , 
\\
\Pr( u, t_{n} \mid u, t_{n-1}) & = 1 - B(t_{n}) \Big\vert_{\smash{B(t_{n-1}) = 0}} \, .  
\end{align*}
Here, vertical bars are used to indicate the initial conditions that should be used when solving Eq.~\eqref{eq:rate-equations} for $t = t_{n}$ and $t' = t_{n-1}$.  

We consider a short-time expansion of the formal solution for $B(t_{n})$ [Eq.~\eqref{eq:formal-solution}] with $t'=t_{n-1}$, where we assume that the time step $\Delta t = t_{n} - t_{n-1} = \const$ is sufficiently small to approximate the integral behind the function $K$ and the corresponding exponential function with a left Riemann sum, namely
\begin{align*}
K(t_{n},t_{n-1}) & \approx [ \koff(t_{n-1}) + \kon(t_{n-1}) ] \Delta t \, ,
\\
\e^{ - K(t_{n},t_{n-1})} & \approx 1 - [ \koff(t_{n-1}) + \kon(t_{n-1}) ] \Delta t \, .  
\end{align*}
The integral term of Eq.~\eqref{eq:bound-population} therefore reduces to
\begin{equation*}
\int_{t_{n-1}}^{t_{n}} \ddf{\tau} \e^{-K(t,\tau)} \kon(\tau) \approx \kon(t_{n-1}) \Delta t \, ,
\end{equation*}
because we only consider terms in $\Delta t$ up to first order.  The relative population in the bound state then takes the form
\begin{equation}\label{eq:approximate-bound-population}
\begin{aligned}
B(t_{n}) & \approx \kon(t_{n-1}) \Delta t + B(t_{n-1})
\\
& \mathrel{\phantom{\approx}} \times [1 - \koff(t_{n-1}) \Delta t - \kon(t_{n-1}) \Delta t ] \, ,
\end{aligned}
\end{equation}
which finally gives rise to Eq.~\eqref{eq:transition-probabilities} in the main text.  Note that Eq.~\eqref{eq:approximate-bound-population} has the exact same form as one would obtain for constant rates and $k_{\text{off/on}} \Delta t \ll 1$.  

Let us now consider the negative logarithm of Eq.~\eqref{eq:conditional-probability}, \emph{i.e.},
\begin{align*}
\mathcal{L}(\vec{\theta} \mid \{ s_{n} \})
& = - \sum_{n=1}^{N} \ln \big( \Pr(s_{n}, t_{n} \mid s_{n-1}, t_{n-1}) \big)
\\
& =  - \sum_{n=1}^{\Nbb} \ln \big( \Pr( b, t_{n}^{b \to b} \mid b, t_{n-1}^{b \to b}) \big)
\\
& \mathrel{\phantom{=}} - \sum_{n=1}^{\Nub} \ln \big( \Pr( b, t_{n}^{u \to b} \mid u, t_{n-1}^{u \to b}) \big)
\\
& \mathrel{\phantom{=}} - \sum_{n=1}^{\Nbu} \ln \big( \Pr( u, t_{n}^{b \to u} \mid b, t_{n-1}^{b \to u}) \big)
\\
& \mathrel{\phantom{=}} - \sum_{n=1}^{\Nuu} \ln \big( \Pr( u, t_{n}^{u \to u} \mid u, t_{n-1}^{u \to u}) \big) \, ,
\end{align*}
where the elements of $\vec{\theta}$ represent the parameters of the model, $t_{n}^{i \to j}$ denotes the time instance of the $n$-th transition from state $i$ to $j$ and $N = \Nbb + \Nub + \Nbu + \Nuu$.  If $k_{\text{off/on}}[F] \Delta t \ll 1$ holds, we can simplify the logarithms of $\Pr(b, t \mid b, t')$ and $\Pr(u, t \mid u, t')$ using the Taylor expansion
\begin{equation*}
\ln(1 - x) = - x + \mathcal{O}(x^{2}) \, ,
\end{equation*}
as $x \to 0$.  With Bell-like rates [Eq.~\eqref{eq:Bell-rates}], the negative log-likelihood then reads (up to some negligible constants)
\begin{equation}\label{eq:neg-log-likelihood}
\begin{aligned}
\mathcal{L} & ( \dxoff, \koff^{0}, \dxon, \kon^{0} \mid \{ s_{n} \})
\\
& = \koff^{0} \Delta t \sum_{n=1}^{\Nbb} \e^{\beta \dxoff \Fbb_{n}} - \Nbu \ln ( \koff^{0} )
\\
& \mathrel{\phantom{=}} + \sum_{n=1}^{\Nub} \beta \dxon \Fub_{n} - \sum_{n=1}^{\Nbu} \beta \dxoff \Fbu_{n} 
\\
& \mathrel{\phantom{=}} + \kon^{0} \Delta t \sum_{n=1}^{\Nuu} \e^{- \beta \dxon \Fuu_{n}} - \Nub \ln ( \kon^{0} ) \, .  
\end{aligned}
\end{equation}
This negative log-likelihood can be minimized with respect to $\koff^{0}$ and $\kon^{0}$, respectively, which gives rise to Eqs.~\eqref{eq:kbu0-eq} and~\eqref{eq:kub0-eq} in the main text.  These can be substituted into the negative log-likelihood expression above, resulting in the following effective two-parameter log-likelihood:
\begin{align*}
\mathcal{L}(\dxoff, & \, \dxon \mid \{ s_{n} \}) = \Nbu \ln \Bigg( \sum_{n=1}^{\Nbb} \e^{\beta \dxoff \Fbb_{n}} \Bigg)
\\
& + \sum_{n=1}^{\Nub} \beta \dxon \Fub_{n} - \sum_{n=1}^{\Nbu} \beta \dxoff \Fbu_{n} 
\\
& + \Nub \ln \Bigg( \sum_{n=1}^{\Nuu} \e^{- \beta \dxon \Fuu_{n}} \Bigg) \, .  
\end{align*}
Again, we have neglected some irrelevant constants.  Furthermore, it turns out that this log-likelihood can be split into two likelihoods, one for each parameter, which can be minimized separately.  These are given by Eqs.~\eqref{eq:dxb-eq} and~\eqref{eq:dxu-eq} in the main text.

\section{Fisher information matrix}\label{app:fim}

By definition, the Fisher information matrix is given by the ensemble-averaged Hessian of the negative log-likelihood [Eq.~\eqref{eq:neg-log-likelihood}].  For the short-time approximated state propagators in the likelihood [Eq.~\eqref{eq:transition-probabilities}], the parameters of the unbinding and rebinding rates are fully uncorrelated, as seen by the fact that
\begin{align*}
\frac{\partial^{2} \mathcal{L}}{\partial \dxoff \partial \dxon} & = \frac{\partial^{2} \mathcal{L}}{\partial \dxoff \partial \kon^{0}} = 0 \, ,
\\
\frac{\partial^{2} \mathcal{L}}{\partial \koff^{0} \partial \dxon} & = \frac{\partial^{2} \mathcal{L}}{\partial \koff^{0} \partial \kon^{0}} = 0 \, ,
\end{align*}
holds for the Bell-like rates of Eq.~\eqref{eq:Bell-rates}.  The $4 \times 4$ Fisher information matrix of the whole system therefore reduces to two separate $2 \times 2$ matrices, namely
\begin{align*}
\mathbf{I}_{\text{off}} = 
\begin{pmatrix}
I_{11}^{\text{off}} & I_{12}^{\text{off}}
\\
I_{12}^{\text{off}} & I_{22}^{\text{off}}
\end{pmatrix} \, , & & 
\mathbf{I}_{\text{on}} = 
\begin{pmatrix}
I_{11}^{\text{on}} & I_{12}^{\text{on}}
\\
I_{12}^{\text{on}} & I_{22}^{\text{on}}
\end{pmatrix} \, ,
\end{align*}
with the following elements:
\begin{align*}
I_{11}^{\text{off}} & = \bigg\langle \frac{\partial^{2} \mathcal{L}}{\partial \dxoff^{2}} \bigg\rangle \, , & I_{11}^{\text{on}} & = \bigg\langle \frac{\partial^{2} \mathcal{L}}{\partial \dxon^{2}} \bigg\rangle \, , 
\\
I_{12}^{\text{off}} & = \bigg\langle \frac{\partial^{2} \mathcal{L}}{\partial \dxoff \partial \koff^{0}} \bigg\rangle \, , & I_{12}^{\text{on}} & = \bigg\langle \frac{\partial^{2} \mathcal{L}}{\partial \dxon \partial \kon^{0}} \bigg\rangle \, ,
\\
I_{22}^{\text{off}} & = \bigg\langle \frac{\partial^{2} \mathcal{L}}{\partial (\koff^{0})^{2}} \bigg\rangle \, , & I_{22}^{\text{on}} & = \bigg\langle \frac{\partial^{2} \mathcal{L}}{\partial (\kon^{0})^{2}} \bigg\rangle \, .  
\end{align*}

The first element associated with the unbinding rate is given by
\begin{equation*}
I_{11}^{\text{off}} = \beta^{2} \koff^{0} \Delta t \sum_{n=1}^{\Nbb} \big\langle \e^{\beta \dxoff \Fbb_{n}} (\Fbb_{n})^{2} \big\rangle \, , 
\end{equation*}
where the calculation of the ensemble average requires a functional form for the distribution of forces $\smash{\Fbb_{n}}$.  Here, we propose a Gaussian with constant variance $\delta F^{2}$, \emph{i.e.},
\begin{equation*}
p[\Fbb_{n}] = \sqrt{\frac{1}{2 \pi \delta F^{2}}} \exp \bigg( - \frac{[\Fbb_{n} - \overline{F}(t_{n}^{b \to b})]^{2}}{\delta F^{2}} \bigg) \, ,
\end{equation*}
with a deterministic protocol $\smash{\overline{F}(t)}$ corresponding to the mean position in the ``bound'' branch at time $t$.  The ensemble average therefore evaluates to
\begin{align*}
\big\langle \e^{\beta \dxoff \Fbb_{n}} & (\Fbb_{n})^{2} \big\rangle = \e^{(\beta \dxoff \delta F)^{2} / 2} \e^{\beta \dxoff \overline{F}(t_{n}^{b \to b})}
\\
& \times \big[ \delta F^{2} + [\overline{F}(t_{n}^{b \to b}) + \beta \dxoff \delta F^{2}]^{2} \big] \, .  
\end{align*}
However, in most practical cases, one can safely replace $\smash{\overline{F}(t_{n}^{b \to b})}$ with $\Fbb_{n}$ without overly skewing the end result.  The elements $I_{12}^{\text{off}}$, $I_{11}^{\text{on}}$ and $I_{12}^{\text{on}}$ can be treated analogously using the same or, in the case of the rebinding elements, a similar force distribution.  

Finally, the inverses of $\smash{\mathbf{I}_{\text{off}}}$ and $\smash{\mathbf{I}_{\text{on}}}$ are given by
\begin{align*}
\mathbf{I}_{\text{off}}^{-1} & = \frac{1}{I_{11}^{\text{off}} I_{22}^{\text{off}} - (I_{12}^{\text{off}})^{2}}
\begin{pmatrix}
I_{22}^{\text{off}} & -I_{12}^{\text{off}}
\\
-I_{12}^{\text{off}} & I_{11}^{\text{off}}
\end{pmatrix} \, , 
\\
\mathbf{I}_{\text{on}}^{-1} & = \frac{1}{I_{11}^{\text{on}} I_{22}^{\text{on}} - (I_{12}^{\text{on}})^{2}}
\begin{pmatrix}
I_{22}^{\text{on}} & -I_{12}^{\text{on}}
\\
-I_{12}^{\text{on}} & I_{11}^{\text{on}}
\end{pmatrix} \, ,
\end{align*}
of which the bounds to the standard errors in the main text [Eq.~\eqref{eq:mle-errors}] can simply be read off.

\section{Synthetic force-extension curves}\label{app:Gillespie-algorithm}

To generate the synthetic force-extension curves used in this paper, we relied on a generalization~\cite{ThanhPriami2015} of the classical Gillespie stochastic simulation algorithm~\cite{Gillespie1977, Gillespie2007} to simulate the exact transition times between the bound and the unbound state.  For our two-state system, the $n$-th transition times $\smash{t^{\text{off/on}}_{n}}$ satisfy the following equations:
\begin{equation}\label{eq:Gillespie-algorithm}
\begin{aligned}
\exp \bigg( - \int_{\ton_{n-1}}^{\toff_{n}} \ddf{\tau} \koff(\tau) \bigg) & = R_{n} \, ,
\\
\exp \bigg( - \int_{\toff_{n}}^{\ton_{n}} \ddf{\tau} \kon(\tau) \bigg) & = R'_{n} \, ,
\end{aligned}
\end{equation}
if the system is initialized in the bound state at $t = \ton_{0}$.  Here, $R_{n}$ and $R'_{n}$ denote uniformly distributed random numbers on $[0,1)$.  For Bell-like rates [Eq.~\eqref{eq:Bell-rates}] and a force protocol
\begin{equation*}
F(t) = 
\begin{cases}
\dot{F} t & s = b~\text{(``bound state'')}
\\
\dot{F} t - F^{*} & s = u~\text{(``unbound state'')}
\end{cases} \, , 
\end{equation*}
Eq.~\eqref{eq:Gillespie-algorithm} can be solved analytically for $\smash{t^{\text{off/on}}_{n}}$, giving
\begin{align}
\notag
\ton_{0} & = 0 \, ,
\\
\label{eq:off-times}
\toff_{n} & = \ln \bigg( \frac{\koff(\ton_{n-1}) - \beta \dxoff \ln(R_{n}) }{\koff^{0}} \bigg) \, , 
\\
\label{eq:on-times}
\ton_{n} & = \ln \bigg( \frac{\kon^{0}}{\kon(\toff_{n}) + \beta \dxon \ln(R'_{n})} \bigg) + \frac{F^{*}}{\dot{F}} \, .
\end{align}
In principle, more complex integrands in Eq.~\eqref{eq:Gillespie-algorithm} can be considered, which arise, \emph{e.g.}, for nonlinear force protocols or more elaborate rate expressions.  In most cases, however, the resulting integrals will not be analytically tractable and Eq.~\eqref{eq:Gillespie-algorithm} has to be solved numerically, which can become computationally demanding.  

The transition times were generated by iteratively evaluating Eqs.~\eqref{eq:off-times} and~\eqref{eq:on-times}, up to the point where no real solution of Eq.~\eqref{eq:on-times} could be found.  Subsequently, the associated time series for a fixed time step $\Delta t$ was constructed by first rounding off the transition times to the accuracy of $\Delta t$, and then calculating the applied force at each time step according to the force protocol $F(t)$.  Instead of immediately breaking off the time series beyond the last transition time $\ton_{n}$, we continued recording the forces in the unbound state for a short time interval to improve the estimates of $\dxon$ and $\kon^{0}$.  

All simulations were conducted using $F^{*} = \SI{6}{\pico \newton}$ and a thermal energy scale of $\beta^{-1} = \SI{4}{\pico \newton \nano \meter}$.  

\end{appendix}

\end{document}